\documentclass[journal]{IEEEtran}
\ifCLASSINFOpdf
\else
\fi
\usepackage{amsmath,amssymb}
\def\forkindep{\mathrel{\raise0.2ex\hbox{\ooalign{\hidewidth$\vert$\hidewidth\cr\raise-0.9ex\hbox{$\smile$}}}}}

\usepackage[colorlinks=true,bookmarks=false,citecolor=blue,urlcolor=blue]{hyperref} %pdflatex

\usepackage{graphicx}
\graphicspath{ {./figures/} }

\usepackage{pifont} % load it first

\begin{document}
%\title{Bare Demo of IEEEtran.cls\\ for IEEE Journals}
%\title{Simoultaneous Deconvolution-Deautocorrelation Image Reconstruction}
%\title{Image Deconvolution and Simultaneous Autocorrelation Inversion}
%\title{Image Restoration via Simoultaneous Deconvolution and Cross-correlation Inversion}
%\title{Image Restoration from Autocorrelations with Simoultaneous Deconvolution}
\title{Deconvolved Image Restoration from Autocorrelations}

\author{Daniele~Ancora*~%\IEEEmembership{Member,~IEEE,}
        %Antonio~Pifferi(?) ~%~\IEEEmembership{Fellow,~OSA,}
%        Somebody~Else(?),~%~\IEEEmembership{Fellow,~OSA,}
        and~Andrea~Bassi~%~\IEEEmembership{Life~Fellow,~IEEE}% <-this % stops a space
\thanks{D. Ancora is with with the Department
of Physics, Politecnico di Milano, 20133 Milan, Italy (*email to: \href{mailto:daniele.ancora@polimi.it}{daniele.ancora@polimi.it}).}% <-this % stops a space
\thanks{A. Bassi is with with the Department of Physics, Politecnico di Milano, 20133 Milan, Italy and with Istituto di Fotonica e Nanotecnologie, Consiglio Nazionale delle Ricerche, Piazza Leonardo da Vinci 32, 20133 Milan, Italy}% <-this % stops a space
\thanks{Manuscript received June XX, 2020; revised June XX, 2020.}}


\markboth{Journal of \LaTeX\ Class Files,~Vol.~14, No.~8, August~2015}%
{Shell \MakeLowercase{\textit{et al.}}: Bare Demo of IEEEtran.cls for IEEE Journals}



\maketitle

\begin{abstract}
%The abstract goes here. The abstract must be between 150-250 words
Recovering a signal from auto-correlations or, equivalently, retrieving the phase linked to a given Fourier modulus, is a wide-spread problem in imaging. 
This problem has been tackled in a number of experimental situations, from optical microscopy to adaptive astronomy, making use of assumptions based on constraints and prior information about the recovered object.
In a similar fashion, deconvolution is another common problem in imaging, in particular within the optical community, allowing high-resolution reconstruction of blurred images.
Here we address the mixed problem of performing the auto-correlation inversion while, at the same time, deconvolving its current estimation.
To this end, we propose an \textit{I}-divergence optimization, driving our formalism into a widely used iterative scheme, inspired by Bayesian-based approaches.
We demonstrate the method recovering the signal from blurred auto-correlations, further
analysing the cases of blurred objects and band-limited Fourier measurements.

%Our method may accept further improvements that can be borrowed from analogous inverse problems, contextualizing this work in a dynamical signal-processing scenario.
\end{abstract}

% Note that keywords are not normally used for peer review papers.

% \fi
\IEEEpeerreviewmaketitle

\section{Introduction}
\IEEEPARstart{D}{econvolution} is an image processing technique commonly used in a number of computer vision and optics applications \cite{richardson1972bayesian,lucy1974iterative}, including astronomy \cite{starck2002deconvolution} and optical microscopy \cite{sibarita2005deconvolution}. The goal of image deconvolution is to restore super-resolved features of an object \cite{mukamel2012statistical} from images acquired with diffraction-limited optics or aberrated wavefronts. 
A different computational imaging technique deals with the estimation of an object's auto-correlation. This is an increasingly studied problem in the fields of not-conventional imaging and far-field diffraction imaging. Applications based on auto-correlation imaging can be found in, but are not limited to, image detection through a turbulent medium (e.g. the atmosphere) \cite{roggemann1996imaging}, lens-less imaging \cite{abbey2011lensless}, hidden imaging \cite{bertolotti2012non} and tomography \cite{ancora2020hidden}. 
In order to form a reconstruction of the object acquired in one of these conditions, typically, one needs to invert the signal's auto-correlation. 
to reconstruct the signal that generated it.
Since the auto-correlation forms a Fourier pair with the modulus of the Fourier transform of the object, the image reconstruction process is often referred to as the Phase-Retrieval (PR) problem.
Furthermore, given that it is not possible to directly measure the phase with camera-sensors, PR compensates for that trying to find the phase connected with the Fourier modulus of the underlined object \cite{fienup2013phase}.
%that is a quantity not possible to be measured with conventional camera-sensors.
There are several approaches to solve this inverse problem: methods based on alternating projections, protocols based on optimization and iterative approaches inspired different inversion strategies.
Although the field is in continuous progress, Shechtman \textit{et al.} provides an extensive review on the topic \cite{shechtman2015phase}.

Measurements relying on the estimation of the object's auto-correlation might be corrupted from blurring, which is given by the limited bandwidth of the detection system or limited by diffraction.
In this case,  two consecutive inverse problems (deconvolution and phase retrieval) should be solved.
Here, instead, we propose an iterative procedure that allows one to obtain a deconvolved image from auto-correlation measurements, solving both problems at the same time. 
We ground the procedure on the \textit{I}-divergence minimization, similarly to Richardson-Lucy \cite{richardson1972bayesian, lucy1974iterative} deconvolution algorithm and in close analogy with Schultz-Snyder approach \cite{schulz1992image}.
For this purpose, it is useful to rearrange the auto-correlation into a convolutional form.
This, renders the problem similar to a deconvolution task in which the target-object is blurred by a kernel that depends on itself.
%Rearranging the auto-correlation into a convolutional form, renders the problem similar to a deconvolution task, but in which the underlined object is blurred by a kernel that depends on itself.
Inspired by blind deconvolution strategies \cite{kundur1996blind}, we discuss how neglecting the kernel's dependence on the object does not alter reconstruction abilities.
%against the complete models and leads to a simple convolutional implementation.
Our approach is similarly developed to established methods in the field of signal processing.
This is a useful design, since the reconstruction can be further improved by adding total-variation regularization \cite{dey2004deconvolution} or deconvolution with unknown kernels \cite{kundur1996blind}.
In the following section, we start introducing the general problem-framework, discussing the context of applicability and different measurement-scenarios.
In sec. \ref{sec:iterativemethod}, we introduce the iterative algorithm that solves both auto-correlation inversion and deconvolution problem. 
Sec. \ref{sec:reconstructionresults} presents the reconstruction results and examines the deconvolution ability of the technique.
Conclusions and potential implementation are discussed in the last part of the article, in sec. \ref{sec:conclusions}.

%demo file is intended to serve as a ``starter file'' for IEEE journal papers produced under \LaTeX\ using IEEEtran.cls version 1.8b and later.
% I wish you the best of success.

\hfill mds

\hfill June 23, 2020

\section{Problem Statement} \label{sec:problemstatement}
First of all, let's define the formalism that we will use throughout the manuscript. 
We will make use of integral formulation, where every equation may be transported into its discrete equivalent by replacing the integral with a summation over the integrated variables.
For the sake of notation, we use one-dimensional functions of the spatial variable $x$, eventually shifted by the quantity $\xi$ due to convolution or correlation operation. 
The formalism can be extended to any dimensionality, and we therefore drop any explicit variable dependence. 
The operator $\mathbf{F}\{.\}$ indicates the Fourier transform and $\cdot$ the element-wise product.
We indicate the cross-correlation between two generic functions $f\left(x \right)$ and $g\left(x \right)$ as:
\begin{equation}
    f \star g = \int \overline{f\left(x \right)} g\left(\xi + x \right) \,dx = \mathbf{F}^{-1}\big\{ \overline{\mathbf{F}\{f\}} \cdot \mathbf{F}\{g\} \big\}
\end{equation}
consequently the auto-correlation of the function $f$ with itself is $f \star f$. The convolution is defined as:
\begin{equation}
    f * g = \int f\left(x \right) g\left(\xi -x \right) \,dx = \mathbf{F}^{-1}\big\{ {\mathbf{F}\{f\}} \cdot \mathbf{F}\{g\} \big\}
\end{equation}
and the auto-convolution as $f*f$.
We are interested in the reconstruction of an unknown object $o$. Its  auto-correlation is defined as:
\begin{equation} \label{eq:autocorrelation_o}
    \chi=o\star o = \mathbf{F}^{-1}\big\{ \| \mathbf{F}\{o\} \|^2 \big\}.
\end{equation}
The identity on the right hand side of eq. \ref{eq:autocorrelation_o}, is given by the power spectral theorem.
Throughout the text we will make use of the greek-pedix $\mu$, which indicates the blurred quantity with added noise, simulating an experimental measurement.
To avoid confusion, from now on we make use of the word "measurement" when referring to a simulated effect that may be measured in a real imaging experiment.
In the present paper we address the case in which the auto-correlation is blurred, a blurred object is used to compute the auto-correlation and the measurement is performed in a band-limited Fourier space.
Although slightly different, it can be seen that these problems are equivalent and can be solved with the same iterative approach.

%There are several problem statements that we aim at addressing with this study, thus in the following we present each version of the problem that we aim at solving, proving that the three statements are equivalent under some specific conditions.

\subsection{Blurred Auto-correlation}\label{sec:blurredautocorr}
\begin{figure}[!t]
\centering
\includegraphics[trim=0 160 0 30, clip, width=3.5in]{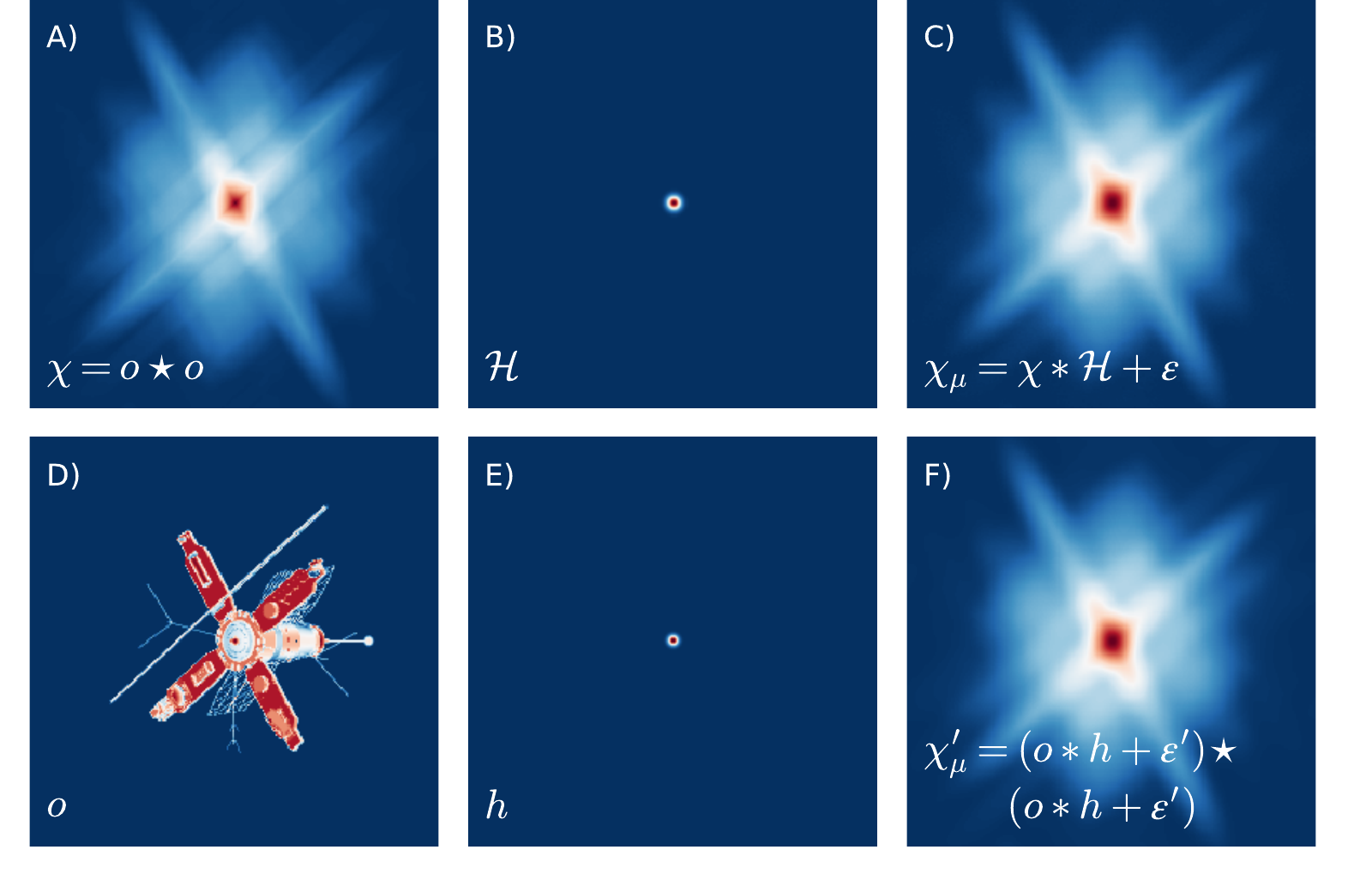}
\includegraphics[trim=0 55 0 375, clip, width=3.5in]{figures/figure01_.pdf}
\caption{Panel A) Noise-free test image $o$ of a satellite \cite{emorySatellite} used to test the reconstruction protocol (size $256 \times 256 \,px$). B) Gaussian kernel $\mathcal{H}$ with standard deviation $\sigma=2\,px$ used to blur the auto-correlation. C) Ideal object's auto-correlation $\chi=o\star o$. D) Auto-correlation $\chi_\mu$ corrupted by Poissonian noise $\varepsilon$ and blurred by $\mathcal{H}$, reproducing the quantitity to which we have access with standard measurements.
All the images are peak-normalized and displayed using a diverging color-map (\textit{RdBu}).
%to emphasize boundaries in the image and the core-part of the auto-correlation.
}
\label{fig:01}
\end{figure}

In some applications, we have access to a statistically computed estimate of the auto-correlation. Looking through turbulent atmosphere in astronomy \cite{roggemann1996imaging}, through scattering slabs or behind corners \cite{bertolotti2012non,katz2014non} and performing hidden tomography \cite{ancora2020hidden} are a few examples.
The light emerging from turbulent environment, in these conditions, has undergone unpredictable scattering events and when detected resembles a random arrangement of intensity distribution. 
It has been proven that under isoplanatic conditions, known also as "memory effect" regime  \cite{freund1988memoryeffect}, the auto-correlation of these patterns shares the same object's auto-correlation.
Thus, the turbidity acts as an opaque auto-correlation lens.
%thus atmosphere and scattering slabs act as an opaque auto-correlation lens.
In order to estimate the auto-correlation of the hidden object, we can process the patterns produced by the light propagation through turbid environment.
Since the auto-correlation is typically averaged through several detections, each of which affected by blurring, the presence of an effective point-spread function (PSF) blurs the final estimate of the auto-correlation.
Thus, we have access to a measurement of the auto-correlation given by:
\begin{equation}
    \chi_\mu = \chi * \mathcal{H} + \varepsilon,
\end{equation}
where $\mathcal{H}$ is a blurring kernel for the original object auto-correlation $\chi=o\star o$. 
Figure \ref{fig:01} shows the test image of a satellite $o$ (panel D) used in this study, and its corresponding auto-correlation $\chi$ (panel A). 
We consider the case in which the image is blurred by a Gaussian kernel $\mathcal{H}$ (panel B) and leads to a noisy measurement of $\chi_\mu$ (panel C). 
Usually these measurements exploit ensemble \cite{freund1990looking} averages thus, for the moment, we consider $\varepsilon$ as a generic form of additive noise.
Approaching this problem would require to deconvolve the auto-correlation first (i.e. using Richardson-Lucy \cite{richardson1972bayesian,lucy1974iterative}) and then to find the signal having that (de-blurred) auto-correlation (e.g. by using Fienup iterative phase retrieval \cite{fienup2013phase}), or vice versa.
%Although this could work in principle, understanding which is the best procedure to pursue for the solution of this problem lies beyond the scope of the manuscript.
%In particular, the image “Satellite” was downloaded from the Emory University image database

Instead, here we want to approach both problems simultaneously.
To do this, we reduce the formula above to the convolutional form $o*\mathcal{K}$, where the object of interest $o$ is subject to the blurring kernel $\mathcal{K}$. 
Let's rearrange the correlations and convolutions in a convenient form:
\begin{align}\label{eq:blurredcorr}
    \chi * \mathcal{H} &= \left(o \star o \right) * \mathcal{H} = o \star \left(o * \mathcal{H} \right) \nonumber \\
    &= o \star \left(\mathcal{H} * o \right) = \left(o \star \mathcal{H} \right) * o \nonumber \\
    &= o * \left(o \star \mathcal{H} \right) \equiv o * \mathcal{K},
\end{align}
where we used the fact that convolution between two functions $f$ and $g$ permutes, $f*g=g*f$, and the correlation-convolution identity $\left(f \star g\right) * h=f \star \left(g*h\right)$. 

In this way, we have introduced a new blurring kernel that depends on the object itself $\mathcal{K}\left[ o \right]= o \star \mathcal{H}$. 
The convolution of the object with $\mathcal{K}$ gives the measured auto-correlation under the effect of a blurring factor $\mathcal{H}$.

% Satellite image: http://www.mathcs.emory.edu/~nagy/RestoreTools/

\subsection{Blurred auto-correlation induced by object-blurring}\label{sec:autocorrblurredobject}
The broad field of Computed tomography (CT) gather together many imaging protocols that let us virtually inspect internal features of the specimen of interest.
Although there exist a number of techniques approaching the problem with different experimental designs, the computational methods used to reinterpret the results are widely shared among them.
In particular, deconvolution and alignment procedures are crucial tasks to obtain faithful reconstructions in CT.
Auto-correlations are centered in the shift-space by definition, thus it is easy to reconstruct a tomographic auto-correlation starting from its projection sequence \cite{ancora2017phase}: the inversion of the latter would return an exact reconstruction of the object with no prior alignment procedure.
Still, this reconstruction is affected by blurring and needs to be treated accordingly, in order to achieve sharp reconstructions.
In alignment-free tomography, we thus often rely on the computation of a blurred version of the auto-correlation $\chi_\mu$ as in:
\begin{equation} \label{eq:autocorrblurredobject1}
    \chi_\mu' = \left(o * h+ \varepsilon' \right) \star \left(o * h+ \varepsilon' \right),
\end{equation}
where the additive noise $\varepsilon'$ perturbs the detection of each blurred projection of the object $o$. 
Dropping the noise term, it is possible to rearrange the problem to reach the same form as eq. \ref{eq:blurredcorr}:
\begin{align} \label{eq:autocorrblurredobject2}
    \left(o * h \right) \star \left(o * h \right) &= \left(o \star o \right) * \left(h \star h \right) \nonumber \\ 
    &= \left(o \star o \right) * \mathcal{H} \nonumber \\
    &= \chi * \mathcal{H},
\end{align}
where the definition of $\mathcal{H} = h \star h$ makes the problem identical to eq. \ref{eq:blurredcorr}. 
In this case, the convolutional kernel becomes $\mathcal{K} = o \star \left(h \star h \right)$.
Equations \ref{eq:autocorrblurredobject1}-\ref{eq:autocorrblurredobject2} imply that the problem is symmetric, thus starting from a blurred auto-correlation it is possible to first de-blur and then de-autocorrelate, or the other way around.
However, as previously stated, we are interested in performing both actions at the same time.
Further details due to the presence of the noise are discussed in the appendix \ref{app:noisetransport}.

\subsection{Blurring due to band-limited Fourier measurement}
Let us consider the case in which we are performing a Fourier measurement limited by a generic window function $W$, as in coherent \cite{miao2011coherent} or partially coherent \cite{clark2012high} diffraction imaging (CDI).
In these experiments, the detector acquires the squared modulus of the Fourier transform of the object, while it is not possible to access the phase information due to electronic limitations.
Given that the phase of an electromagnetic signal is not measurable, imaging is usually performed by solving a phase retrieval problem \cite{shechtman2015phase} having the Fourier modulus as fixed constraint.
The signal detected under these conditions can be expressed as:
\begin{equation}
    \mathcal{M}_\mu = \| \mathbf{F}\{o\} \|^2 \cdot W + \varepsilon'',
\end{equation}
where the window function is due to the limited size of the detector, which inevitably cuts some high spatial frequencies.
If we compute the auto-correlation based on the sole measurement of $\mathcal{M}_\mu$, we have that:
\begin{equation}\label{eq:bandlimmeasure}
    \chi_\mu'' = \mathbf{F}^{-1}\{\mathcal{M}_\mu\},
\end{equation}
where we used the Wiener-Kinchin power spectral theorem.
We could again reach a formulation similar to eq. \ref{eq:blurredcorr} by making use of the convolution theorem:
\begin{align}
    \mathbf{F}^{-1}\{\| \mathbf{F}\{o\} \|^2 \cdot W \} &= \mathbf{F}^{-1}\{ \mathbf{F}\{ \chi \} \cdot \mathbf{F}\{ \mathcal{H} \} \} \nonumber \\
    &= \mathbf{F}^{-1} \{ \mathbf{F}\{\chi * \mathcal{H} \} \} \nonumber \\
    &= \chi * \mathcal{H}.
\end{align}
Where, for the band-limited Fourier measurements, the convolution kernel reduces to $\mathcal{K} = o  \star \, \mathbf{F}^{-1}\{W\}$.
This result for the auto-correlation is analogous to the convolutional blurring in a band-limited measurement, as dictated by the limited aperture of a detection objective:
%This result for the auto-correlation is in analogy with the fact that convolutional blurring always implies a band-limited measurement, that in optical imaging is dictated by the limited size of the objective used to collect the light: 
a convolutional kernel is a filter in the frequency domain.
So far, we considered the effect of the noise to be negligible, but a thorough discussion on this can be found in the appendix \ref{app:noisetransport}.

%\IEEEpubidadjcol

\section{Iterative Method}\label{sec:iterativemethod}
As discussed in the previous section, several measurements fall within the same class of inverse problem, that involve the inversion of the auto-correlation coupled with the deconvolution of the reconstruction at the same time.
Similarly to de-autoconvolution procedure \cite{choi2005iterative}, our problem can be viewed as a generalized convolution with a kernel containing the object itself, correlated with a factor $\mathcal{H}$. 
In our formalism, $\mathcal{H}$ contains the description of the problem that we are interested in, among the three classes examined in section \ref{sec:problemstatement}.
Before stepping further, let's introduce an approximation by dropping the explicit functional dependence on $o$ from $\mathcal{K}\left[o\right] \xrightarrow{} \mathcal{K}$. 
This approximation implies the complete knowledge of the convolution kernel and we will motivate this choice in the following paragraph.

Let us call a generic measurable distance between $\chi_\mu$ and $\chi^*$ as $d\left(\chi_\mu || \chi^* \right)$, where $\chi^*=o^* * \mathcal{K}$ is the best estimate of the auto-correlation. 
Provided that $\chi_\mu$ is the measurement, the optimal object's reconstruction $o^*$ is reached when such distance $d\left(\chi_\mu || \chi^* \right)$ is minimized.
Thus, the reconstruction problem turns into:
\begin{equation} \label{eq:argmin}
    o^* = \arg \min d\left(\chi_\mu || \chi^* \right).
\end{equation}
A necessary (but not sufficient) condition to satisfy the equation above 
%eq. \ref{eq:argmin} 
is given by setting its functional derivative, with respect the object, to be optimized $o^*$ to zero:
\begin{equation}\label{eq:minimumtozero}
    \frac{\delta}{\delta o^*} d\left(\chi_\mu || \chi^* \right) = 0.
\end{equation}

\begin{figure*}[t]
\centering
\includegraphics[trim=20 100 115 90, clip, width=7in]{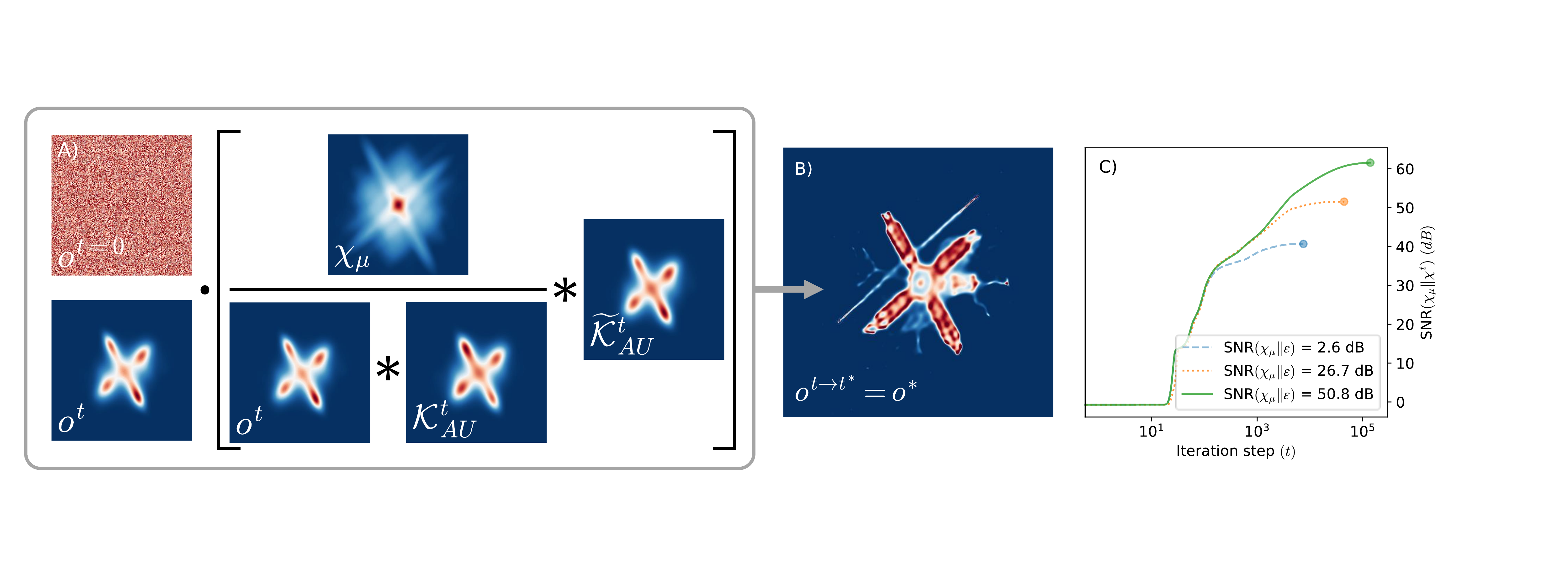}
% where an .eps filename suffix will be assumed under latex, 
% and a .pdf suffix will be assumed for pdflatex; or what has been declared
% via \DeclareGraphicsExtensions.
\caption{Panel A) presents a schematic of the AU iteration, as in eq. \ref{eq:anchorUpdate2}. A random guess is chosen as starting point $t=0$, and then the results after $t=60$ iterations are displayed. 
Panel B) visualizes the result after around $10^5$ iterations of the deblurred auto-correlation inversion subject to Poissonian noise $\lambda = 2^4$. This reconstruction refers to section \ref{sec:blurredautocorrResults}.
In C), the plot relative to the SNR during the iteration steps for all the noises considered.}
\label{fig:02}
\end{figure*}

There are several possible choices for $d\left(\chi_\mu || \chi^* \right)$, each of which would turn the reconstruction search into a different inverse problem. 
We deal with intensity measurements and it is therefore desirable that $o\left(x\right) \geq 0, \forall x$. To fulfill this requirement we chose the Csisz\'ar's \textit{I}-divergence \cite{csiszar2004information} as distance estimator:
\begin{align}\label{eq:Idivergence}
    & I\left(\chi_\mu || \chi^* \right) =  \\ 
    & \quad = \int \bigg\{ \chi_\mu\left(\xi\right) \ln \left[ \frac{\chi_\mu\left(\xi\right)}{\chi^*\left(\xi\right)} \right] + \chi^*\left(\xi\right) -\chi_\mu\left(\xi\right) \bigg\} \,d\xi. \nonumber
\end{align}
Such quantity is a generalization of the relative entropy, known as Kullback-Leibler distance.
This measure is common in iterative convolution/correlation inversion schemes \cite{schulz1992image,choi2005iterative}, whereas choosing the Euclidean norm $d\left(\chi_\mu || \chi^* \right) = \| \chi_\mu - \chi^* \|^2$ would usually turn the reconstruction into a least-square problem (see \cite{starck2002deconvolution} for a thorough overview).
Setting its first derivative to zero as in eq.\ref{eq:minimumtozero}:
\begin{align}
    \frac{\delta}{\delta o^*} I\left(\chi_\mu || \chi^* \right) = \frac{\delta}{\delta o^*} I\left(\chi_\mu || o^* * \mathcal{K} \right)
\end{align}
corresponds to calculate:
\begin{align}
    \frac{\delta}{\delta o^*} I\left(\chi_\mu || \chi^* \right) &= - \frac{\delta}{\delta o^*} \int \Big\{ \chi_\mu \log \left[ \chi^*\left( \xi \right) \right] - \chi^*\left( \xi \right) \Big\}  \,d\xi \nonumber \\
    &= \int \left(1- \frac{\chi_\mu}{\chi^*\left( \xi \right)} \right) \frac{\delta}{\delta o^*}\left[ \chi^*\left( \xi \right) \right] \,d\xi.
\end{align}
Setting it to zero and inserting the functional derivative as calculated in the appendix \ref{app:partialderivative} (eq. \ref{eq:finalfunctionalderivative}) we obtain:
\begin{equation}\label{eq:integraltozero_partial}
    \int \left(1- \frac{\chi_\mu}{\chi^*\left( \xi \right)} \right) \mathcal{K}(\xi-x) \,d\xi = 0.
\end{equation}
A closed solution does not exist in general, however, an efficient scheme to solve it is provided by the Picard iteration \cite{isaacson2012analysis}.
This implies the use of a fixed-point iterative formulation:
\begin{equation} \label{eq:fixedpointiteration}
    o^{t+1}(x) = o^t (x) \lambda^t(x),
\end{equation}
where the updating rule $\lambda^t(x)$ is:
\begin{equation}\label{eq:updateintegral}
    \lambda^t(x) = \int \frac{\chi_\mu\left(\xi\right)}{\int o^{t}\left(x\right) \mathcal{K}\left(\xi - x\right) \,dx}\mathcal{K}\left(\xi-x\right) \,d\xi.
\end{equation}
This is similar to the expectation-maximization approach often used in similar procedures \cite{shepp1982maximum}.
The usage of Csisz\'ar's I-divergence guarantees that eq. \ref{eq:fixedpointiteration} converges to its minimum subject to non-negative constraints for the solution $o$. 
Further details and properties of this approach were extensively analyzed in seminal works \cite{snyder1992deblurring, shepp1982maximum, choi2005iterative}.

It is worth noting that Eq. \ref{eq:updateintegral} is rigorous when $\mathcal{K}$ is given as a fixed prior and closely relates to RL-like schemes. However, similarly to the case of blind deconvolution \cite{chaudhuri2016blind}, we do not have access to an exact estimate for $\mathcal{K}$. In the latter, the point spread function (PSF) estimate is updated via RL-steps, using the current object estimate as the (transiently fixed) kernel that blurs the recorded image \cite{kundur1996blind}. 
Here, instead, we decide to anchor the update (AU) of this kernel with the current estimate of the object, inferring a refined kernel at each step of the form:
\begin{equation}
    \mathcal{K}_{AU} \left(\xi\right) = \int \overline{o^{t} \left(x\right)} \mathcal{H}\left(x+\xi \right) \,dx.
\end{equation}
In a compact, convolutional notation, the whole iterative AU-process becomes:
\begin{align}
    \mathcal{K}^{t}_{AU} &= o^t \star \mathcal{H} \label{eq:anchorUpdate} \\
    o^{t+1} &= o^{t} \left[ \left(\frac{\chi_\mu}{o^{t} * \mathcal{K}^{t}_{AU}}\right)*\widetilde{\mathcal{K}}^{t}_{AU}\right] \label{eq:anchorUpdate2}
\end{align}
where $\widetilde{\mathcal{K}} = \mathcal{K}\left(-x\right)$ denotes reversed axis. 
At each iteration the convolutional kernel is refined, taking into account the previous object estimate, and used to de-blur the auto-correlation into the new object reconstruction $o^{t+1}$.
In the particular case $\mathcal{H}=\delta$, the problem is not reduced to the deautoconvolution method, as one would think \cite{choi2005iterative}, since $\mathcal{K}_\delta = o \star \delta = {\overline{o\left(-x\right)}}$.
For ease of notation, we do not use normalization constants and we consider all the quantities to be normalized by their total intensity, prior to processing. 
This procedure guarantees the conservation of the total energy throughout the minimization.

\section{Reconstruction Results}\label{sec:reconstructionresults}
In order to test the reconstruction protocol, let's define a metric to assess the quality of the images obtained.
Given a generic measured signal corrupted by additive noise, $s_\mu = s+\varepsilon$, we estimate the fidelity of the measurement against the noiseless signal with the signal-to-noise ratio (SNR):
\begin{align}
    \text{SNR}\left(s_\mu \| s \right) &= 20 \, \log_{10}{\left[\frac{\int s_\mu\,dx}{\int \left(s_\mu-s\right) \,dx}\right] } \label{eq:SNR1} \\
    \text{SNR}\left(s_\mu \| \varepsilon \right) &=20 \, \log_{10}{\left[\frac{\int s_\mu\,dx}{\int \varepsilon \,dx}\right] }. \label{eq:SNR2}
\end{align}
These equivalent forms are commonly used for intensity detection and expressed in decibel, $dB$.
In the cases considered in this work, $s$ (and $s_\mu$) might stand for either the auto-correlation $\chi$ (and $\chi_\mu$) or the object $o$ (and $o_\mu$).

\subsection{Reconstruction from a blurred auto-correlation}\label{sec:blurredautocorrResults}
We start our analysis discussing the results of the AU on the satellite image subject to the experimental conditions described in section \ref{sec:blurredautocorr}.
Thus, we want to reconstruct the object $o^*$ from the auto-correlation $\chi_\mu = \chi * \mathcal{H}+\varepsilon$.
%in Fig. \ref{fig:01}A.
%although the protocol was tested using also different images.
We start with $\chi$ having 16 bit accuracy, to which we add a random Poissonian noise with different parameters $\lambda=2^{12}$, $2^8$ and $2^4$.
We recall that $\lambda$ in a Poisson distribution corresponds either to the mean value and to the variance of the values generated.
The effect of the noise introduced to the measured signal is evaluated in terms of its SNR as in eq. \ref{eq:SNR2}.
Here, we consider a Gaussian point-spread function $h$ with standard deviation $\sigma=2 \,px$, defining the blurring kernel for the auto-correlation as $\mathcal{H} = h*h$, which is also Gaussian with a broader $\sigma=4 \,px$.
%This was done to be able to compare these results with those of Fig. \ref{fig:03}, where we consider the experiment described in section \ref{sec:autocorrblurredobject}.
Adding Poissonian noise (with the $\lambda$ defined above) to the blurred object auto-correlation, $\chi * \mathcal{H}$, results into three different SNR$\left(\chi_\mu \| \varepsilon \right)=2.6\,dB, \,  26.7\,dB$ and $50.8\,dB$ for the $\chi_\mu$.
%SNR$\left(\chi_\mu \| \varepsilon \right)=34.5, 82.7$ and $106.8$ for the $\chi_\mu$.
For the post-processing, we subtract the noise mean to the auto-correlation and then we take its absolute value,
since negative correlation would be unphysical for the reconstruction of an intensity image. 
After this, we normalize the the latter by dividing for its total intensity.
As from the problem statement, we know the generic blurring function $\mathcal{H}$ and we start from a strictly positive initial guess of the object $o^{t=0}=o^i$, energy-normalized so that $\int o^t(x) \,dx = 1$.
In principle, $o^i$ could accommodate any known prior, such as sparsity information or a low-resolution estimate of the object, if available.
However, we assume the object to be completely unknown, feeding a random initial guess to the AU-iterations.
Fig. \ref{fig:02} presents a view of the iterative scheme at $t=60$ in the case of low-noise, where it appears evident the formation of a blurred version of the satellite object.
%in eq. \ref{eq:anchorUpdate}-\ref{eq:anchorUpdate2}.
Since AU minimizes the \textit{I}-divergence,
%given in eq. \ref{eq:Idivergence}
we rely on the signal-to-noise ratio to monitor the reconstruction quality as the iteration progresses.
%Euclidean distance would be a good choice, but 
We compute the SNR$\left(\chi_\mu\| \chi^t \right)$ as defined in eq. \ref{eq:SNR1}, treating the current estimate for the (blurred) auto-correlation $\chi^t=o^t*\mathcal{K}_{AU}^t$ as a perturbation with respect the given quantity $\chi_\mu$.
Thus, the noise of the current auto-correlation estimate at $t$-step would be $\varepsilon^t=\chi_\mu - \chi^t$. 
%between the measured $\chi_\mu$ and the current auto-correlation estimate $\chi^t=o^t*\mathcal{K}_{AU}^t$.
While progressing, we expect that our method would increase the SNR until an optimal point is reached, after which the reconstruction quality might degrade.
%We point out again that this measurement does not refer to the direct object reconstruction against the object, but it is measured on the auto-correlation sequences.
%With this measurement scheme, the noise term $\varepsilon$ is given by the distance be
%As an example, Fig. \ref{fig:01}D shows the measured $\chi_\mu$ with the lowest SNR$\left(\chi_\mu \| \varepsilon \right)=34.5$.
%The schematic of the AU iterations and the results of the algorithm are presented in Fig. \ref{fig:02}.
Therefore, we stop the iteration when the SNR of the reconstruction results lower then the previous value.
The plot in Fig. \ref{fig:02}C shows the trend for the three noises considered and the dots represents the stopping point for each reconstruction.
The initial growth is similar for all cases but, in general, low-noise measurement leads to higher reconstruction SNR.
%However, the plot in Fig. \ref{fig:02}C shows an almost constant SNR increase starting from around $10^2$ steps.
%For this study, we limit the total number of iteration to $10^6$ steps that returns nearly identical reconstructions for the three noise level considered as the one shown in panel B.
%Fine details in the image (i.e. the thin antennas) were correctly recovered, in particular in the high SNR-case.
%In fact, the SNR followed a similar trend in the three cases considered, where only the noisiest measurement begins to saturate after $10^5$ steps.
This growth can be compared with the measured SNR$\left(\chi_\mu \| \varepsilon\right)$, since in either case it is calculated directly on auto-correlations.
%For the noisiest situations, the final SNR is higher than the originally measured one, as if the noise effect would have been compensated during the problem inversion.
%For the intermediate noise level, in $10^6$ we achieved the same SNR as the original measurement while for the highest SNR we did not reach such level.
%The three final outputs in Fig. \ref{fig:02}A,B,C are visually similar, with fine details in the image (i.e. the thin antennas) better recovered for the highest SNR measurement.

\begin{figure}[t!]
\centering
\includegraphics[trim=0 10 0 10, clip, width=3.5in]{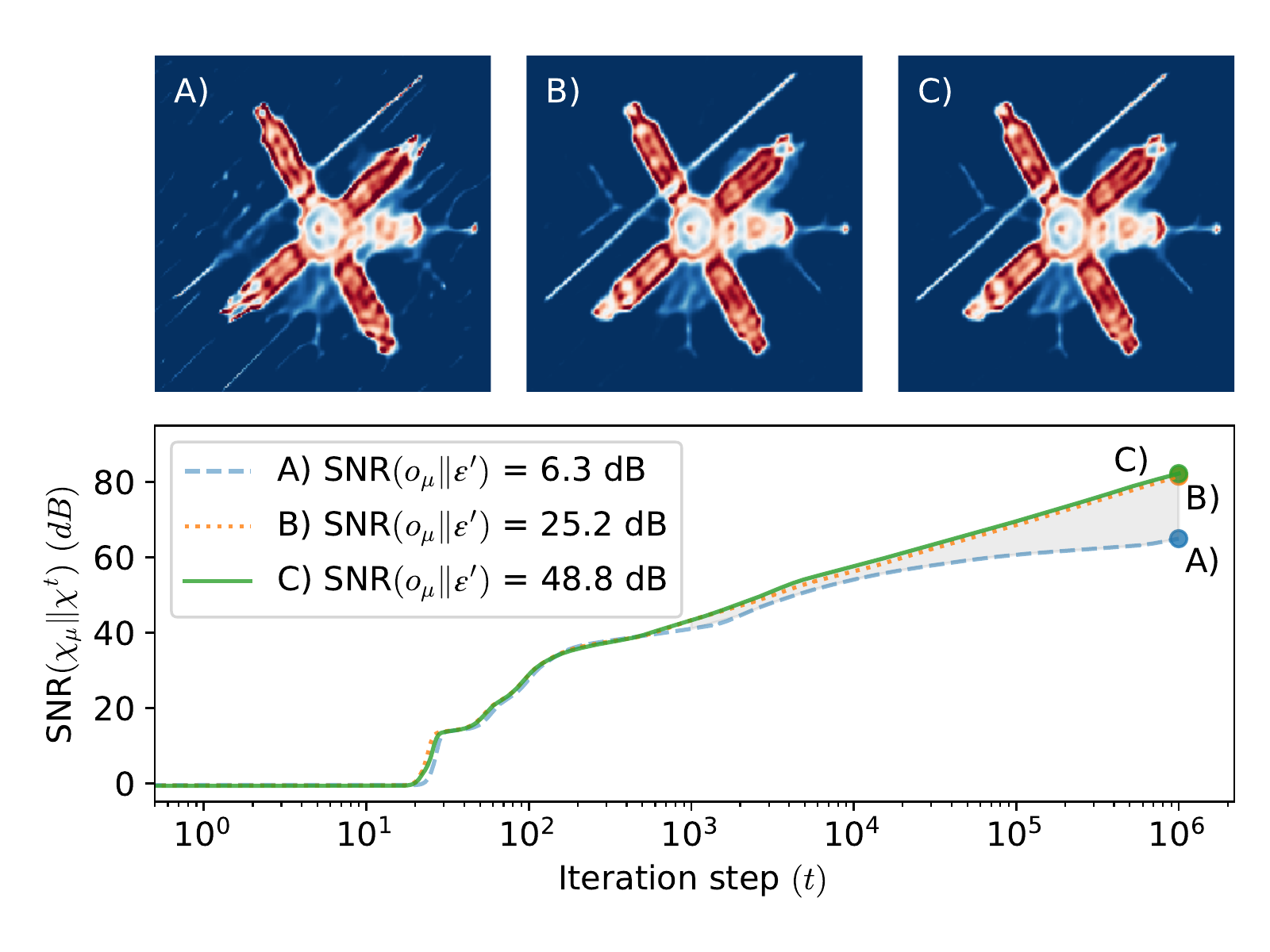}
% where an .eps filename suffix will be assumed under latex, 
% and a .pdf suffix will be assumed for pdflatex; or what has been declared
% via \DeclareGraphicsExtensions.
\caption{Simulation results for the experiment described in section \ref{sec:autocorrblurredobject}. A) Output of the reconstruction after $10^6$ iteration for the lowest SNR. B) Result for the intermediate SNR and C) for the highest. The plot underneath visualized the SNR trend as a function of the iteration step (in log-scale).}
\label{fig:03}
\end{figure}

\subsection{Reconstruction from an auto-correlation of a blurred object}\label{sec:autocorrblurredobjectResults}
Here we discuss the results in case the auto-correlation is obtained by a blurred estimate of the object, as in $\chi_\mu = \left(o*h+\varepsilon'\right)\star\left(o*h+\varepsilon'\right)$.
This part refers to section \ref{sec:autocorrblurredobject} and we consider $o$ acquired with 16 bit accuracy.
%In this case, the auto-correlation is obtained by a blurred estimate of the object as in $\chi_\mu = \left(o*h+\varepsilon'\right)\star\left(o*h+\varepsilon'\right)$.
%are shown in Fig. \ref{fig:03},. 
The kernel $h$ was chosen to be Gaussian with standard deviation $\sigma=2\,px$, and the blurred object is perturbed by the addition of Poisson's noise with three different $\lambda=2^{12}$, $2^8$ and $2^4$, as in the previous case.
Here the SNR of the measurement is referred to the measured object $o_\mu = o * {h}+\varepsilon'$, thus SNR$\left(o_\mu \| \varepsilon' \right)$, rather than to its auto-correlation.
%Since the test image is sparse (there is an extended zero-region around the satellite), the SNRs span a wide range of values.
%In particular, the image perturbed with $\lambda=2^{12}$ has a low SNR$=6.3\,dB$, whereas the others have higher signal-to-noise ratio as SNR$=25.2\,dB$ and $48.8\,dB$.
Remarkably, in all cases, the AU-iteration converged to faithful reconstructions during an entire run of $10^6$ iterations. 
The results are shown in Fig. \ref{fig:03}.
Panel A, displays the output result with SNR$=6.3\,dB$, whereas panel B refers to SNR$=25.2\,dB$ and panel C to SNR$=48.8\,dB$.
%but similar image quality was achieved already at about $10^5$ iterations.
%From the plot below, we notice that the SNR saturates to a stable value and the reconstruction does not improve any further.
As expected, the result improves reducing the noise and fine details become visible for either the reconstructions in panels \ref{fig:03}B and \ref{fig:03}C.
In this case, the quantity SNR$\left(o_\mu \| \varepsilon \right)$ cannot be directly compared with the values shown on plot underneath. 
The former is referred to the object, not to the auto-correlation.
We should compare, instead, the SNR$\left(\chi_\mu \| \chi*\mathcal{H} \right)$ that turns to be 
%$62.6\,dB$, $86.1\,dB$ and $94.5\,dB$ 
$32.9\,dB$, $44.5\,dB$ and $56.6\,dB$ 
respectively for panels \ref{fig:03}A, \ref{fig:03}B and \ref{fig:03}C.
%Even if the noise may have strong effects in the direct object measurement, transported into the auto-correlation turns into high SNR for the auto-correlation.
In this case, a seemingly good SNR$\left(\chi_\mu \| \chi*\mathcal{H} \right)=32.9\,dB$ gives poor reconstruction quality (Fig. \ref{fig:03}A), especially if compared to the one obtained in the previous section.
This may be due to the fact that additive noise in this experimental regime is transported into the auto-correlation in a non-trivial fashion, as discussed in the appendix \ref{app:noisetransport}.
%Here, we could not reach SNR higher than the measured ones for the auto-correlations, even if longer runs may lead to improved reconstruction for B and C.
However, these results are visually comparable to those obtained in the previous section.

\begin{figure}[t]
\centering
\includegraphics[trim=0 10 0 10, clip, width=3.5in]{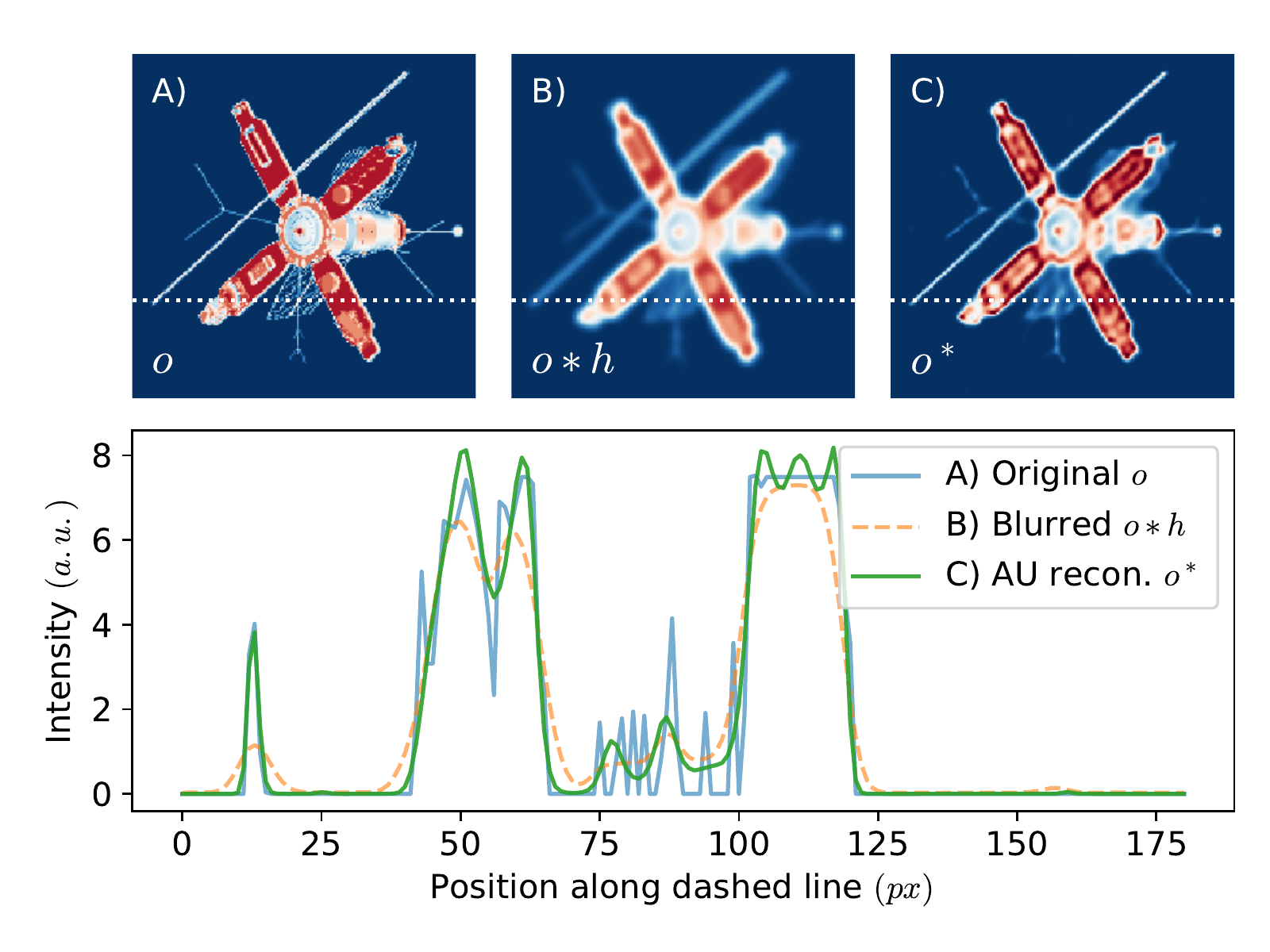}
% where an .eps filename suffix will be assumed under latex, 
% and a .pdf suffix will be assumed for pdflatex; or what has been declared
% via \DeclareGraphicsExtensions.
\caption{Above plot, profile comparison between the original object, shown in panel A), the blurred object B) as in $o*h+\varepsilon$ and the results of the AU algorithm C) for the experiment described in \ref{sec:autocorrblurredobject}. The dashed lines in A,B,C represent the profile analyzed along the images. The noise was generated using $\lambda=2^8$.}
\label{fig:04}
\end{figure}

It is worth noting that the problem treated here is equivalent to a classical deconvolution approach.
To understand this, we focus on the dashed line profiles in Fig. \ref{fig:04}.
The original object $o$ in panel A is compared against its blurred version $o_\mu = o * h + \varepsilon'$ (panel B) and with our reconstruction (panel C).
The latter was aligned to the original object by finding the peak of their cross-correlation and translating the reconstruction accordingly.
The results can be interpreted by focusing the analysis on the leftmost spike in the plot of Fig. \ref{fig:04}.
We notice that the smooth-peak (orange dashed) is sharpened by the AU iterations (green line), approaching the original object's resolution (blue line, reference).
Along this profile, the recovered intensity exceeded the expected value in the central region (over-shoot): this might happen also with RL-deconvolution, where the overall energy it is preserved, but local fluctuations may alter the reconstructed values. 
The same applies for all the features in the image, as can be seen by a more accurate analysis of the plot.
In particular, the flat region on the plot's right side exhibits oscillations (under- and over-shoot) around the correct value.
These results suggest that AU is inverting an auto-correlation sequence, deblurring the result at the same time.
However, our protocol it is not meant to be another deconvolution scheme.
%does not want to be another deconvolution scheme.
With single direct acquisitions of blurred objects, standard deconvolutions are still the best options to follow.
On the other hand, one could encounter situations in which the object is not visible as a whole \cite{ancora2018optical, preibisch2014efficient} and the sample needs to be measured from different perspectives. 
This, typically requires an alignment or co-registration procedure for of each view, in order to properly form the reconstruction.
Here, protocols based on auto-correlation inversion may lead to alternative reconstruction strategies, since auto-correlations are inherently aligned to the shift-space origin.

\section{Conclusion}\label{sec:conclusions}
So far, we have discussed the problem of simultaneous deconvolution and auto-correlation inversion. 
Our approach was grounded on the theoretical basis of \textit{I}-divergence optimization: the results from RL-deconvolution \cite{richardson1972bayesian, lucy1974iterative}, de-autocorrelation \cite{schulz1992image} and de-autoconvolution \cite{choi2005iterative} strongly inspired the development of present algorithm.
Reconstruction using auto-correlation that are directly blurred, calculated from a blurred object or obtained with a band-limited detection belong to the same class of inverse problem.
Here, we studied how they can be described by the same formalism, proposing and testing the AU-algorithm as reconstruction scheme of a commonly used test image from an auto-correlation sequence.
Remarkably, the method shows promising performances under a variety of experimental noise-levels, leading to faithful and robust reconstructions.
Since the problem is similar to a phase retrieval process (with added deblurring), we expected the solution to be statistically sensitive to prior initialization \cite{shechtman2015phase}.
As a matter of fact, the method behaved well in all the tested conditions, converging to visually similar reconstructions even with independent random initialization.
It is worth noting that, being the auto-correlation insensitive to the absolute positioning of the object, any independent reconstruction grows randomly positioned in space.
This is considered a trivial non-uniqueness connected with this class of inverse problems, as it is an inherent feature of the auto-correlation (and to the Fourier modulus of the object).
Even if not explicitly mentioned, every reconstruction was realigned with the original object for visual comparison only.
The object absolute position is encoded in its Fourier phase, thus sharing the same class of shift-invariant solutions with the latter \cite{shechtman2015phase}.
Although interesting per se, this approach might have potential applications where ensemble averages are the sole (or the optimal) way allowed to estimate the object's auto-correlation.
As discussed, imaging through turbulence \cite{roggemann1996imaging}, hidden imaging \cite{bertolotti2012non} (or related problems) are scenarios where AU could be applied.
In particular, our method may allow further developments in the field of alignment-free multi-view reconstruction, since the auto-correlation is insensitive to translations.
This can be exploited promoting the reconstruction in the shift-domain and then bring back the imaging procedure to the measurement space, via auto-correlation inversion \cite{ancora2017phase}.
Of course, a similar scheme has to deal with the solution of a generic phase retrieval problem, that is well known to produce unstable results under various experimental conditions \cite{fienup2013phase}.
This is why we decided to approach the problem using as a fix-point iteration scheme, rather than including deconvolution in alternate-projection PR methods. 
%such as the hybrid input-output phase retrieval.
Having chosen to optimize the \textit{I}-divergence gives also room for further improvements by, for example, importing strategies developed in standardized deconvolution methods.
Among the others, blind kernel updates \cite{chaudhuri2016blind} and total-variation regularization \cite{dey2004deconvolution} are two features that we are considering to include in future developments of the AU protocol.

%By choosing this approach, our procedure can be tuned similarly to iterative deconvolution methods based on Bayesian approaches, by for example including blind kernel updates \cite{chaudhuri2016blind}, total variation regularization \cite{dey2004deconvolution}, or eventually extended to approach different kind of distance estimation \cite{starck2002deconvolution}.

%\subsubsection{Subsubsection Heading Here}
%Subsubsection text here.

% or


\appendices
\section{Noise in cross-correlation problems} \label{app:noisetransport}
Since we have used the Csisz\'ar's \textit{I}-divergence, the ideal scenario for the inversion would be when the signal is corrupted by Poisson noise. 
However, it has been proven that similar approaches well behave in the case of high SNR measurements with experimental noise.
Here we discuss how the noise in all the cases can be related to the additive noise described in section \ref{sec:blurredautocorr}, where we had that $\chi_\mu = \chi*\mathcal{H}+\varepsilon$.
Separating all the noise-terms in $\chi'_\mu$ leads to:
\begin{align}
    \chi'_\mu &= \left(o * h+ \varepsilon' \right) \star \left(o * h+ \varepsilon'
    \right) \nonumber \\
    &= \chi*\mathcal{H} + \left(o*h\right) \star \varepsilon' + \varepsilon' \star \left(o*h\right) + \varepsilon' \star \varepsilon'.
\end{align}
We can recognize that this expression coincides with $\chi_\mu$ when considering:
\begin{equation}
    \varepsilon = \left(o*h\right) \star \varepsilon' + \varepsilon' \star \left(o*h\right) + \varepsilon' \star \varepsilon'.
\end{equation}
For uncorrelated random noise with uniform variance throughout the image, the last term is simply $\varepsilon' \star \varepsilon'=\delta$. 
Furthermore, if the noise is not correlated with the signal $o*h$ itself, also the first two terms vanish when $\varepsilon'\xrightarrow{}0$ 
For the band-limited measurement considered in eq. \ref{eq:bandlimmeasure}, it is easy to recognize that the noise is related to $\varepsilon$ via:
\begin{equation}
    \varepsilon = \mathbf{F}^{-1}\{\varepsilon'' \}.
\end{equation}
The formulation above implies that, for random high-frequency additive noise, $\varepsilon$ perturbs the auto-correlation as a strongly peaked function $\varepsilon\xrightarrow[]{}\delta$, centered at the zero-shift space of the auto-correlation.
In all the aforementioned cases, the noise does not diverge when computing the auto-correlations.

\section{Partial variational derivative of $\chi^*$} \label{app:partialderivative}
Here we consider an operative case in which we ignore the explicit dependence of $\mathcal{K}\left[o\right]$ from $o$. This implies that we are able to estimate $\mathcal{K}$ as not being a functional form of $o$.
As for the definition of variational derivative \cite{Engel2011}, we let the function $o$ variate by a small amount $o + \epsilon \theta$, thus we compute:
\begin{equation}
    \frac{d}{d\epsilon} F \left[ o + \epsilon \theta \right] \bigg|_{\epsilon=0} =: \int \frac{\delta F\left[ o \right]}{\delta \left[ o\left(x \right) \right]} \theta\left(x \right) \,dx
\end{equation}
If we apply this to the optimal auto-correlation $\chi^*$ that minimizes the \textit{I}-divergence and dropping the asterisk in $o^*$, we have:
\begin{align}
    \frac{d}{d\epsilon} \chi^* \left[ o + \epsilon \theta \right] \bigg|_{\epsilon=0} &= \frac{d}{d\epsilon} \int \left[ \overline{o(x)} + \epsilon\overline{\theta(x)} \right] \mathcal{K}\left(x-\xi \right) \bigg|_{\epsilon=0} \,dx \nonumber \\
    &= \frac{d}{d\epsilon} \int \epsilon \theta\left(x\right) \mathcal{K}\left(x-\xi \right) \bigg|_{\epsilon=0} \,dx \nonumber \\
    &= \int \theta\left(x\right) \mathcal{K}\left(x-\xi \right) \,dx.
\end{align}
%Since in the present section there will be no ambiguities in its definition, let us drop the asterisk in $o^*$ that denotes the optimal estimation that we aim at obtaining.
Sometimes, it is useful to identify the action of the variational derivative on a Dirac's delta function, that would let us to re-write:
\begin{equation} \label{eq:finalfunctionalderivative}
    \frac{\delta}{\delta \left[ o\left(x \right) \right]}\left[ o * \mathcal{K}\right] = \delta * \mathcal{K} = \mathcal{K}\left(\xi - x\right).
\end{equation}
Let us notice that this is similar to the functional derivatives that in Richardson-Lucy deconvolution leads to a convolution with a reversed kernel \cite{richardson1972bayesian,lucy1974iterative}, where the kernel remains stable during the whole minimization.

\section{Complete variational derivative of $\chi^*$}
This appendix aims at giving the complete variational derivative of the auto-correlation.
Again, let us drop the asterisk in $o^*$ denoting the optimal object's estimation while calculating the complete derivative of the optimal auto-correlation $\chi^*$.
To accomplish this task, let us consider that all the signals we are dealing with are real, thus we don't treat explicitly complex conjugates. 
We have that:
\begin{align}
    \frac{\delta}{\delta \left[ o\left(x \right) \right]}\left[ o * \mathcal{K}\left[o\right]\right] = \left( \frac{\delta}{\delta \left[ o\left(x \right) \right]} o \right)* \mathcal{K}\left[o\right] + \nonumber \\ 
    + o * \left( \frac{\delta}{\delta \left[ o\left(x \right) \right]} \mathcal{K}\left[o \right] \right).
\end{align}
The first term is identical to what obtained in the appendix \ref{app:partialderivative}, the second, instead:
\begin{align}
    \frac{\delta}{\delta \left[ o\left(x \right) \right]} \mathcal{K}\left[o \right] &=  \frac{\delta}{\delta \left[ o\left(x \right) \right]} \left[ o \star \mathcal{H} \right] \nonumber \\
    &= \left( \frac{\delta}{\delta \left[ o\left(x \right) \right]} o \right) \star \mathcal{H} = \delta \star \mathcal{H}.
\end{align}
Here, the terms' order is important, since the correlation operator is not commutative. We have that:
\begin{align}
    \frac{\delta}{\delta \left[ o\left(x \right) \right]}\left[ o * \mathcal{K}\left[o\right]\right] &= \delta * \mathcal{K} + o * \left(\delta \star \mathcal{H}\right) \nonumber \\
    &= \delta * \mathcal{K} + \left(\delta \star \mathcal{H}\right) * o \nonumber \\
    &= \delta * \mathcal{K} + \delta \star \left(\mathcal{H} * o \right) \label{eq:completederiv}
\end{align}
We have kept the Dirac delta functions to exploit correlation/convolution properties and reach the formulation above.
Eq. \ref{eq:completederiv} suggests that the correct update for the iterative method is:
\begin{align}
    o^{t+1} &= o^{t} \left[ \left(\frac{\chi_\mu}{o^{t} * \mathcal{K}^{t}}\right)*\widetilde{\mathcal{K}}^{t} + \left(\frac{\chi_\mu}{o^{t} * \mathcal{K}^{t}}\right) \star \left(o * \mathcal{H}\right) \right] \\
    &= o^t \left[ \lambda_1^t \left(x\right) +\lambda_2^t \left(x\right) \right]. \label{eq:completemodel}
\end{align}
The update rule $\lambda^t \left(x\right)$ is now composed by two terms, in analogy with what found for the auto-correlation inversion by Schultz and Snyder in \cite{schulz1992image}.
The first term $\lambda_1^t \left(x\right)$ can be recognized as the formulation given by the AU iteration in eq. \ref{eq:anchorUpdate2}, instead, $\lambda_2^t \left(x\right)$ is a new term that enforces the update.
We can test how much the result changes in case we consider: A) the full model given in eq. \ref{eq:completemodel}, B) the AU iteration (obtained by setting $\lambda_2^t \left(x\right)=0$) and C) using the remaining term only (where $\lambda_1^t \left(x\right)=0$).

\begin{figure}[!t]
\centering
\includegraphics[trim=0 10 0 10, clip, width=3.5in]{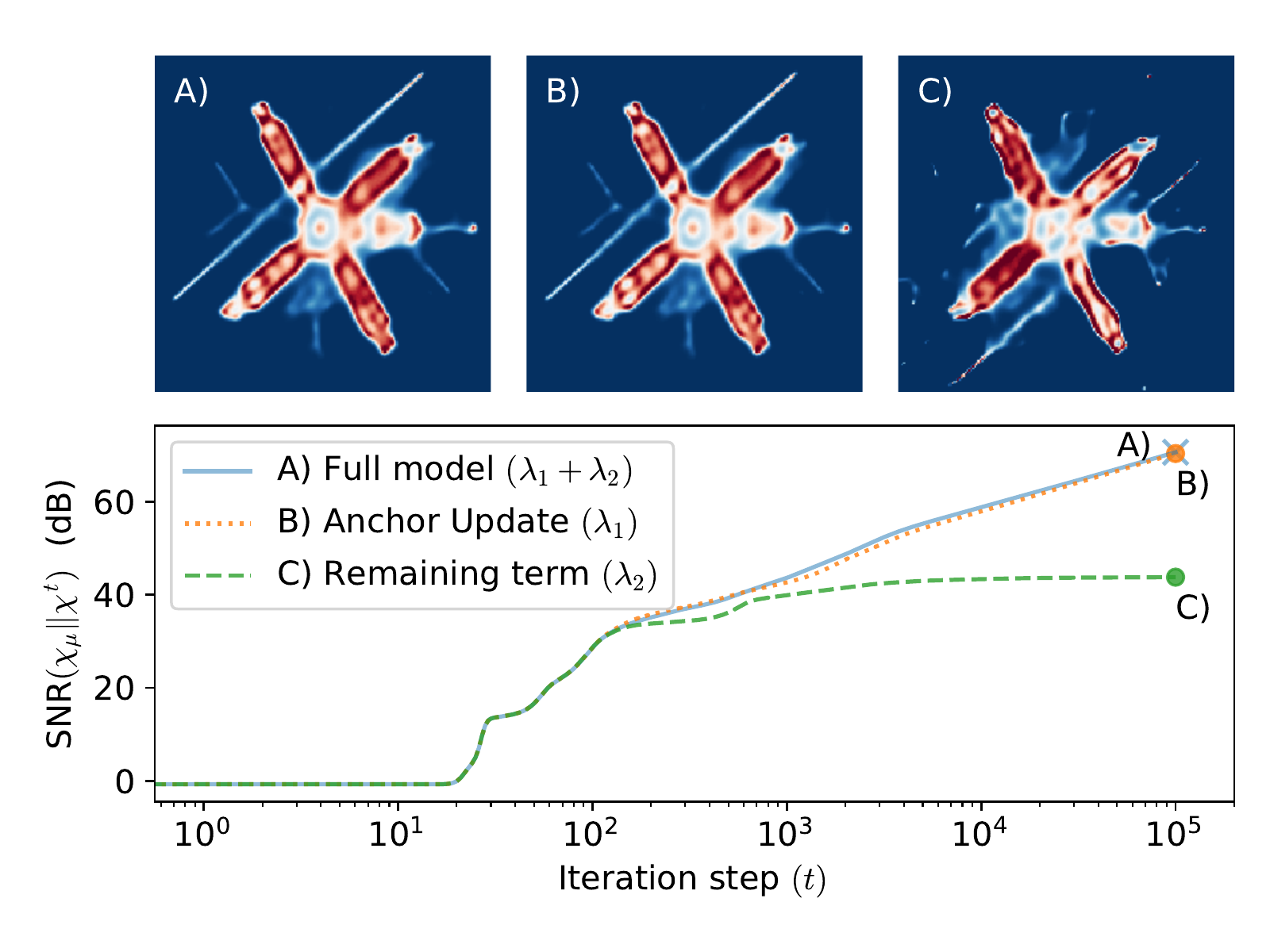}
% where an .eps filename suffix will be assumed under latex, 
% and a .pdf suffix will be assumed for pdflatex; or what has been declared
% via \DeclareGraphicsExtensions.
\caption{Result comparison of the full-model (updating with $\lambda_1^t + \lambda_2^t$) A), the AU approximation (update $\lambda_1^t$) B) and the remaining term (update $\lambda_2^t$) C) as described in eq. \ref{eq:completemodel}. The plot underneath shows that making use of the AU approximation does not degrade the image quality achievable by the model whereas, instead, the $\lambda_2^t$ does not suffice reconstruction purposes.}
\label{fig:05}
\end{figure}

In figure \ref{fig:05}, we report the results of the comparison carried on the numerical experiment described at \ref{sec:blurredautocorr}.
From the plot in the lower part, we notice that the full model completes the run with the highest SNR and the AU closely performs, following a very similar trend.
On the other hand, the remaining term started departing from an ideal behaviour after $10^2$ iterations, slowly saturating the SNR of the reconstruction.
The output images of the full model in Fig. \ref{fig:05}A and the AU in Fig. \ref{fig:05}B are almost indistinguishable, whereas the remaining term returned a noisier image Fig. \ref{fig:05}C.
In the latter, we notice that the longest antenna in the image is formed both on the right and on the wrong side.
This suggests that this method may suffer from twin-image reconstruction problem, were the algorithm stagnates into a reconstruction given by the combination of the actual object $o$ and its mirrored version $\widetilde{o}$.
This is a well-known problem in the original phase retrieval implementation, the error-reduction method, that stimulated further studies to overcome this algorithmic behavior \cite{fienup1986phase}).
%Finally, for this numerical experiment we have run $10^5$ iterations, to show how already at this point the object's reconstruction closely resembles the expected one.

%\section{}

\section*{Acknowledgment}
%European Union's Marie Skłodowska-Curie Actions (HI-PHRET, g.a. 799230) and H2020 Laserlab Europe V (g.a. 871124).
H2020 Marie Skłodowska-Curie Actions (HI-PHRET project, 799230); H2020 Laserlab Europe V (871124).
The authors thank Prof. Antonio Pifferi for scientific and logistic support and Gianmaria Calisesi for inspiring discussions.

\ifCLASSOPTIONcaptionsoff
  \newpage
\fi



\bibliographystyle{IEEEtran}

% Bibliography
\bibliography{IEEEabrv, bibtex/bib/reference2.bib}

%\begin{thebibliography}{1}
%\bibliography{references.bib}


%\end{thebibliography}


\vfill 
\begin{IEEEbiography}[{\includegraphics[width=1in,height=1.25in,clip,keepaspectratio]{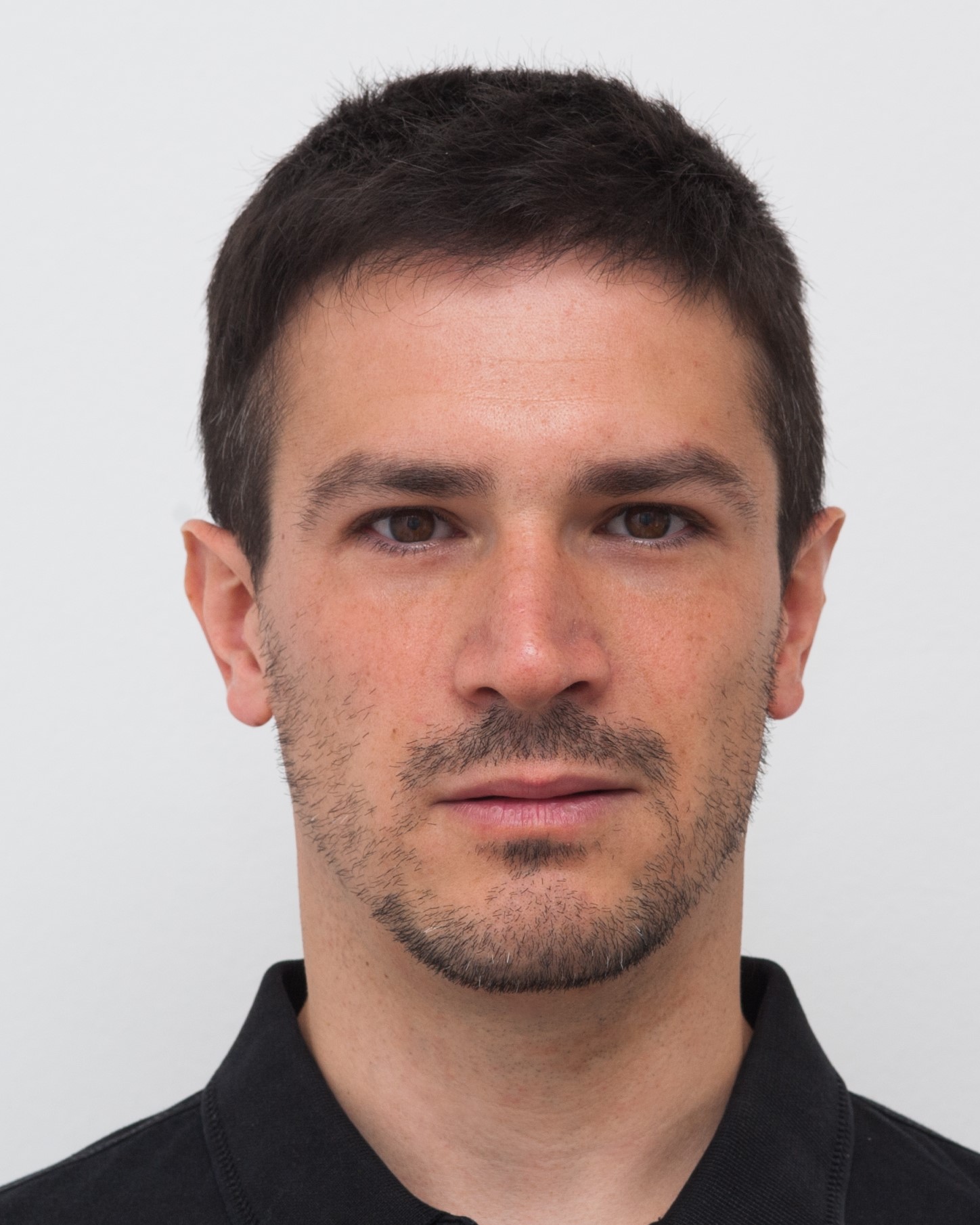}}]{Daniele Ancora}
received his B.S. and M.S. at the Department of Physics of La Sapienza Università di Roma. He completed his Ph.D. studies at the Department of Material Science and Technology at the University of Crete. During this period he was a Marie Curie Early Stage Researcher (MSCA-ITN, g.a.: 317526) at the Foundation for Reseearch and Technology Hellas (IESL-FORTH).
He was Post-Doc at Institute of Nanotechnology at Consiglio Nazionale delle Ricerche (CNR-NANOTEC) in Rome.
He is currently running a Marie Curie Individual Fellowship (MSCA-IF, g.a.: 799230) as a Post-Doc at the Department of Physics of Politecnico di Milano.
His research interests span from disordered photonics to biomedical optics in diffusive regime at the interface between computational, theoretical and experimental imaging.  
\end{IEEEbiography}

\begin{IEEEbiography}[{\includegraphics[width=1in,height=1.25in,clip,keepaspectratio]{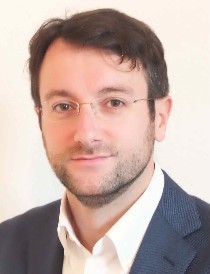}}]{Andrea Bassi} 
received his Ph.D. in Physics at Politecnico di Milano in 2006. He conducted research at Beckman Laser Institute, University of California (Irvine) in 2005/2006 as Research Specialist, at Politecnico di Milano from 2009 to 2014 as tenured Researcher, at Max Planck Institute of Molecular Cell Biology and Genetics (Dresden) in 2013/2014 as Marie Curie Fellow.
Currently he is an Associate Professor at the Department of Physics of Politecnico di Milano. His scientific interests include optical tomography and microscopy for biological and preclinical applications.
\end{IEEEbiography}

%\newpage



%\vfill


\end{document}